\documentclass[10pt,twocolumn]{article}
\usepackage{fullpage}
\usepackage{color}
\usepackage{amsmath}
\usepackage{graphicx}

\newcommand{\bR}{{\bf R}}

\newcommand{\bx}{{\bf x}}
\newcommand{\hilbert}{{\cal H}}

\def\bra#1{\mathinner{\langle{#1}|}}
\def\ket#1{\mathinner{|{#1}\rangle}}

\begin{document}

\title{Discovering correlated fermions using quantum Monte Carlo}
\author{Lucas K. Wagner and David M. Ceperley}

\maketitle
\begin{abstract}
It has become increasingly feasible to use quantum Monte Carlo (QMC) methods to study correlated fermion systems for realistic Hamiltonians. 
We give a summary of these techniques targeted at researchers in the field of correlated electrons, focusing on the fundamentals, capabilities, and current status of this technique.
The QMC methods often offer the highest accuracy solutions available for systems in the continuum, and, since they address the many-body problem directly, the simulations can be analyzed to obtain insight into the nature of correlated quantum behavior.
\end{abstract}

\section{Introduction}

Because of their interactions, correlated fermions offer unique properties that are not possible otherwise.
In materials, these properties include magnetism, high temperature superconductivity, large magnetoresistance, and many other effects.
Actually, interactions are important in all materials--without interactions, the $s$, $p$ and $d$ orbitals in an atom would be degenerate.
However, in many materials, these interaction effects can be folded into effective single-body pictures.
For the example of the atomic levels, it is a decent approximation to set the effective energy of the $2p$ orbitals higher than the $2s$ orbitals, even though the difference is an interaction effect.
In order to predict these effective single-body pictures, it is worth developing quantum techniques to incorporate the effects of interactions and correlations from first principles.
But correlated electron systems are different; there may be no effective single-body model that describes the physics well, and so many-body quantum techniques are even more important to obtain a qualitative understanding of the material. 

For condensed matter systems, the program to solve for the behavior of ensembles of quantum particles was laid down in the early part of the 20th century.
That program is quantum mechanics, which is most compactly encapsulated in the many-body Schr\"odinger equation.
The challenge in bringing this program to completion is that the Sch\"odinger equation increases in dimension with the number of particles, making interacting systems intractable for more than a few particles. 
The basic mathematical problem to be solved has not really changed in almost a century.

As we shall discuss, the level of understanding of many body systems and the degree to which they can be solved mathematically has progressed dramatically in the intervening years. 
Just as large numerical calculations have been able to provide insight into ensembles of classical particles, climate systems, and supernovae, faster computers and better algorithms have enabled much better calculations of quantum many-body systems.
The important quality that runs through the applications listed here is that since the QMC techniques directly simulate the correlations and directly work with the many-body wave function, the (approximate) solutions can be analyzed to obtain qualitative physical information about correlated fermions.
In this review, we will consider a few examples of correlated fermion systems where quantum Monte Carlo techniques have resulted in new understanding.
These examples run from the pure model of interacting fermions, the electron gas, to superconducting atoms, to interacting electrons in highly realistic models of materials.
In all these examples, the direct simulation of many-body effects leads to both higher accuracy and improved qualitative information about the physics in the material.

\section{Solving the Schr\"odinger equation using Monte Carlo}
\label{monte_carlo}

\begin{figure}
\includegraphics{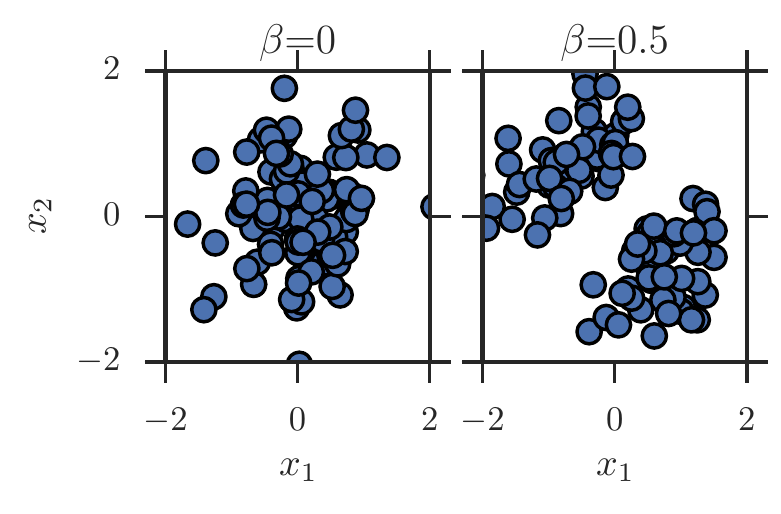}
\caption{Sampling a two-particle function $\rho(\bx)=\exp\left( -0.5 (x_1^2+x_2^2) - \beta/(x_1-x_2)^2\right)$ using Monte Carlo techniques.}	
\label{fig:sample_demo}
\end{figure}

There are a number of articles that summarize the details of quantum Monte Carlo techniques\cite{foulkes_quantum_2001,ceperley_path_1995}.
In this article, we seek to give the interested researcher some idea of how to interpret a quantum Monte Carlo calculation and whether or not such a calculation might be necessary or useful to analyze a physical problem.

In many-body quantum problems, the Monte Carlo method is usually used to perform high dimensional integrals. 
The advantage of Monte Carlo is that high dimensional integrals can be evaluated efficiently. 
This is done by {\it sampling} values of a high dimensional coordinate $\bx$.
As an example, consider Fig~\ref{fig:sample_demo}.
For the interacting distribution, the particles avoid one another, which is observed directly in the samples; when $x_1$ is near zero, $x_2$ is pushed to either side.
The important point here is that while one could represent the density of samples using grids in two dimensions, the sampling approach scales much better when the number of dimensions is increased.
This property allows Monte Carlo processes to access many body systems consisting of thousands of particles in 3D, which corresponds to sampling $\bx$ with multiple thousands of dimensions.
The reason this technique works well is similar to the reason that polls can obtain accurate estimates of an entire population from a small subsection: application of the central limit theorem.

When evaluating an integral in Monte Carlo, one should break the integrand down into two parts: the probability distribution $\rho$ and the quantity to be averaged $A$.
$\rho$ must be non-negative everywhere and must integrate to 1.
If $\rho$ can be sampled efficiently, then the integral
\begin{equation}
\int{A(\bx)\rho(\bx) d\bx}
\end{equation}
can be evaluated by generating random variables $\bx_i$ with probability density $\rho(\bx)$ and averaging the values of $A(\bx_i)$.
The variance of $A$ when sampled according to $\rho$ must be finite so that the central limit theorem applies. 
Once those criteria are satisfied, the Monte Carlo integration can be performed.
Technical improvements in the method often consist of sampling $\rho$ more efficiently and choosing $\rho$ and $A$ such the variance of $A$ is as small as possible.

The major Monte Carlo techniques can be understood in this paradigm.
Variational Monte Carlo (VMC) generates $\rho$ proportional to the trial wave function squared, which allows the evaluation of the properties of that wave function.
The energy can be optimized with respect to a parameterization to implement the variational method.
Diffusion Monte Carlo (DMC) instead generates $\rho$ proportional to the so-called mixed distribution, which allows access to the ground state properties.
Path Integral Monte Carlo (PIMC) generates $\rho$ proportional to the finite-temperature density matrix, which allows access to finite temperature properties.

\subsection{Approximation techniques: Fixed node}

PIMC and DMC are particularly interesting techniques because in the limit of infinite sampling they are exact, they attain the exact thermal or ground state properties.
However, for many fermion Hamiltonians, the variance of $A$ increases exponentially with the size of the system because $A$ has a rapidly fluctuating sign, which gives this the name `the sign problem.'
For some Hamiltonians, including the homogeneous electron gas\cite{C015}, the sign problem can be overcome for systems with enough particles to be useful.
For some other Hamiltonians, such as half-filled Hubbard models on the square lattice, $\rho$ can be chosen such that there is no sign problem.
Unfortunately, most Hamiltonians do have a sign problem and so approximation techniques are necessary to perform useful calculations.

For ground state calculations using diffusion Monte Carlo, the most common approximation is the fixed-node approximation\cite{C015}.
In this approximation, the zeros of a trial wave function $\Psi_T$ are assumed to be the same as the exact ground state.
The resulting energy is an upper bound to the exact ground state energy, so multiple different $\Psi_T$'s can be tried to find the best solution.
A very similar approach exists for the finite temperature PIMC techniques\cite{C069}. For systems for which the wave function is necessarily complex, a generalization, the fixed-phase method\cite{C080} is used.

The fixed-node approximation also allows one to calculate the properties of excited states by preparing a trial wave function with the appropriate symmetry.
Under certain conditions\cite{foulkes_symmetry_1999}, the fixed-node approximation for an excited state is also an upper bound to its energy.
This can be used to estimate gaps and to calculate metal-insulator transitions.
Even when it's not clear if the conditions of Ref~\cite{foulkes_symmetry_1999} apply, this technique seems to work well in practice.

Another promising approach, not covered here in detail and in general less well-tested than fixed node diffusion Monte Carlo, is to walk in the Slater determinant space, known as auxilliary field QMC, or AFQMC.
This technique has a different approximation\cite{zhang_quantum_2003}, which may lead to higher accuracy than fixed node calculations\cite{zhang14}.

\subsection{Calculating quantities}

Since quantum Monte Carlo techniques work directly with the many-body wave function or density matrix, one can evaluate many properties that can be written as an expectation value of the wave function.
This is simply performing an integral using Monte Carlo as explained earlier in this section, with associated $A$ and $\rho$. 
Similar considerations also apply; $A$ must have finite variance. It can sometimes be the case that a particular expectation value is difficult or impossible to evaluate, even if the wave function is known.
Some examples of this kind of operator include the one-particle Green's function and the many-electron polarization operator\cite{souza_polarization_2000} in 3D\cite{hine_localization_2007}. 

The fact that the efficiency of Monte Carlo techniques depends mainly on the variance also works in its favor.
Consider that for a chemical system, the total energy may be of order 1,000-10,000 Hartrees (a Hartree is $\sim$27 eV). 
Properties of interest are often at energy scales close to 10-100 meV, or roughly 1-10 mHartree, which corresponds to factors of  10$^{5}$ to 10$^{7}$ between the size of the total energy and the size of the effect.

So how can a Monte Carlo method possibly resolve quantities so precisely? 
The answer is the zero-variance property. 
For energy, the quantity averaged is the local energy $E_L(\bR)=\Psi(\bR)^{-1}\hat{H}\Psi(\bR)$. 
For an eigenstate, $E_L(\bR)$ is a constant, and thus its variance is zero.
For $\Psi(\bR)$ close to an eigenstate, the variance of $E_L(\bR)$ becomes quite small. 
For example, on recent calculations of the cuprates, while the total energy was around 3,000 Hartrees, the standard deviation of the local energy was only about 4.5 Hartrees. 
To obtain statistical error bars of around 2 mHartree, one thus needs to generate around 5 million independent samples, a formidable task, but certainly attainable with modern resources.

\section{Homogeneous electron gas}

The homogeneous electron gas (HEG) is one of the basic models for condensed matter physics. It consists of an infinite system of interacting electrons (in 1D, 2D or 3D) at a fixed density, parameterized by $r_s$ (defined in 3D by $\rho_e=3/(4\pi r_s^3)$  where $\rho_e$ is the electron number density) with a uniform static background of opposite charge to provide charge neutrality. Originally proposed as a model of a simple metal such as sodium, it became more prominent with the advent of density functional theory (DFT); the correlation energy of the HEG is at the kernel of all DFTs. The calculation of the HEG energy is a problem that QMC is ideally suited for since there is a simple Hamiltonian leading to relatively simple wave functions, a simple  HF state and no core electrons or electron-ion interactions to worry about.  Its calculation\cite{C015} was the first fermion DMC calculation and still the most cited.   The use of accurate QMC energies within DFT did much to advance the popularity and standardization of DFT for electronic structure calculations.

The HEG at high density ($r_s<1$) approaches non-interacting fermions; perturbation methods can estimate its properties. In the density range of $1<r_s<5$ the electrons are moderately correlated and become strongly correlation for $r_s>5$.   It was Wigner that conjectured that the HEG would form a crystal at low density. This phase is now known as the Wigner crystal. A Wigner crystal is the purest form of strong correlation; the formation of the insulating localized phase only depends on the electron-electron interaction, not on band effects, or even fermion statistics. While Wigner supposed that the transition would occur when kinetic energy was on the order of potential energy ( by definition $r_s=1$), QMC calculations find that it happens at a million times lower density, at $r_s=100$!

QMC work since the 1980's on the HEG has extended this early work.   There has been exploration of more accurate trial wave functions, for example with backflow correlations\cite{C125} and advances in methodology such as finite size scaling\cite{C157,C182}. There has been calculation of the response of the HEG to weak electric fields\cite{C092}, of its momentum distribution\cite{C222}, its quasi-particle strength  and its effective mass\cite{C204}. Recent Path Integral Monte Carlo calculations\cite{C237,C242} for non-zero temperature have provided tabulations of how the correlation energy of the HEG changes with density and temperature.  Finally, there have been studies of the magnetism in the Wigner crystal \cite{C178}.

However, the HEG does not contain all of the physics relevant to materials and there is no direct experimental system on which to validate the QMC methods. In the next section we discuss helium, which though not an electronic system, can help validate the method.

\section{$^3$He: a strongly correlated atomic system} \label{sec:helium3}

In contrast to the HEG which lacks a direct experimental realization, experiments on helium are very clean and precise. At ambient pressures and low temperatures one can consider a helium atom an elementary particle: because the first electronic excitation is 10$^5$ K, for temperatures when quantum effects of the atoms are relevant (on the order of 1K), the probability of an electronic excitation is on the order of $\exp(-10^5)$.  The Born-Oppenheimer interaction between the atoms is relatively well known, both experimentally and computationally, much better than between any other types of atoms because the polarizability of helium atoms is small.  Because the interaction between atoms is weak, the many-body ground state is a quantum liquid, unique in the periodic table. The naturally occurring isotope, $^4$He is a boson. Below 2.1K it makes a transition to a bose-condensed superfluid. The transition and properties are calculated using Path Integral Monte Carlo \cite{C095}. The other stable isotope $^3$He is a spin $\frac{1}{2}$ fermion.  

Liquid $^3$He is the simplest strongly correlated fermion system that is experimentally accessible.  In 3D, at low temperatures, $^3$He forms a Fermi liquid and was discovered to become a p-wave superfluid at  temperatures about 1000 times lower than the fermi energy, at a mK scale \cite{leggett2006}. Because the length scales in the superfluid are so large, and energy scales are so small, QMC simulations have not been able to access this superfluid state. It remains a challenge for the future.  However, description of the Fermi liquid state has improved over the years, so that computational errors (finite-size and fixed-node) are of the order of 0.1K/atom \cite{IterativeBackflow}; small compared to the kinetic energy of 20K /atom.

What we want to discuss is the magnetism in the solid $^3$He phase; QMC has played a very important role in understanding the physics. On pressurizing liquid $^3$He to 30 bars, it forms a b.c.c. solid. On cooling down to mK temperatures, the spins form a N\'{e}el state consisting of 2 planes of up-spins followed by 2 planes of down spins, the so-called uudd state. A nearest neighbor Heisenberg models for a b.c.c. lattice would order into an antiferromagnetic state; this more complicated symmetry is unexpected.

\begin{figure}
\includegraphics[width=\columnwidth,height=.8in]{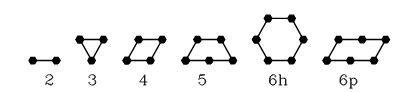}
\caption{Ring exchanges considered for a 2D lattice.}
\label{fig:desexch}
\end{figure}

Thouless in the 1960's \cite{thouless65}, based on earlier work of Herring, formulated an exchange model for understanding the magnetic properties of a quantum crystal. One can think of the atoms as spin $\frac{1}{2}$ hard spheres. In a crystal, the atoms spend much of their times vibrating around fixed lattice sites.   Very rarely, there is a multi-atom ring exchange: two or more atoms exchange positions; it is because of this exchange that the spin of the atoms, and the fact that they are fermions, comes into play. Implicit in the ring exchange model is that the energies of interstitials or vacancies have a much higher energy and do not occur at low temperature. If even 0.1\% of lattice sites were vacant, then the exchange caused by vacancy motion would dominate the magnetic properties.  It is the rate of the rare ring exchanges that determines the magnetic state of the solid. If we call $p$ a particular exchange, such as nearest neighbor two-body exchange,  then it can be shown that the low energy Hamiltonian is given by  $H = \sum_p J_p  P_{\sigma}$ where $J_p$ is the tunneling rate and $ P_{\sigma}$ is an operator which causes a ring exchange of spins.  Examples of exchanges for the 2D triangular lattice are illustrated in Fig. \ref{fig:desexch}. The sum is over all possible ring exchanges. \cite{roger11}

The computation of the magnetic properties is now reduced to two problems. First, how to determine the $J_p$'s and second, how to solve the resulting Hamiltonian.  In turns out that QMC methods are quite capable of determining $J_p$.  The rate can be mapped onto a classical problem: what is the classical free energy to make such a ring exchange?  We have developed specialized QMC techniques\cite{C039} to do such calculations that work even if the rates are exponentially small and have performed such calculations for solid $^3$He and for the Wigner crystal.  The second problem is more difficult in general because the resulting Hamiltonian is frustrated and would have a QMC sign problem. The most effective technique is to perform an exact diagonalization of the Heisenberg Hamiltonian; this is possible for systems of up to $\sim$50 spins by using all of the lattice symmetries \cite{PhysRevB.60.1064}.  One can also calculate properties using expansions and perturbations from high temperature or high magnetic field.

One finds something a little surprising: the resulting Hamiltonian is not dominated by a  particular ring exchange as you might expect for tunneling processes, but many different ring exchanges contribute.  To understand anything at all about the experimental situation, one needs a model that includes 2, 3 and 4 particle ring exchanges.   This is called the multi-spin Hamiltonian. It is the competition between these various exchanges that gives rise to the frustrated uudd state. To achieve quantitative agreement between experiment and the ab initio derived exchanges, one needs to consider exchanges of 6 atoms\cite{C226}.  The situation for 2D helium crystals \cite{PhysRevB.60.1064} is more complicated but also interesting, both from experiment and from computation.   Whether such a situation of multi-spin exchanges applies to strongly correlated electronic situation is an important question. But we learn that the simplest models (nearest neighbor Heisenberg) are not necessarily the correct ones. Constructing such models by fitting experimental data, is suspect. Section 7 will discuss recent attempts to perform such ``downfolding'' on electronic systems.

\section{Quantum Monte Carlo for chemical systems}

The underlying non-relativistic theory of condensed matter physics is the first principles Hamiltonian:
\begin{align}
\hat{H} &= - \hbar \sum_i \frac{\nabla_i^2}{2m_e} 
- \hbar \sum_\alpha \frac{\nabla_\alpha^2}{2m_\alpha}
- \sum_{i\alpha}\frac{Z_\alpha e^2}{4\pi\epsilon_0 r_{i\alpha}} \notag \\
&+ \sum_{ij}\frac{e^2}{4\pi\epsilon_0 r_{ij}} 
+ \sum_{\alpha\beta} \frac{Z_\alpha Z_\beta e^2}{4\pi\epsilon_0 r_{\alpha\beta}} .
\label{eqn:se}
\end{align}
In practice, most calculations are performed with a few approximations of this Hamiltonian.
First, the nuclei are usually fixed, which ignores the kinetic energy of the nuclei.
Second, pseudopotentials are typically used to remove the core electrons.
There are now pseudopotentials\cite{burkatzki_energy-consistent_2007,burkatzki_energy-consistent_2008,trail05} that are designed for the high accuracy requirements of quantum Monte Carlo.

\subsection{Trial wave functions}
\label{sec:trial_wavefunctions}
\begin{table*}
\caption{Wave functions commonly used in quantum Monte Carlo calculations.}
\label{table:trial_wf}
\begin{tabular}{|c|c|c|}
\hline
Wave function & Size extensive & Cost \\
\hline
Slater-Jastrow (SJ) & Yes & ${\cal O}(N_e^3)$ \\
Multi-Slater-Jastrow (MSJ) & No & ${\cal O}(N_e^{3-4})$ \\
Backflow-Jastrow & Yes & ${\cal O}(N_e^4)$ \\
AGP/Pfaffian-Jastrow & Difficult & ${\cal O}(N_e^3)$ \\ \hline
\end{tabular}
\end{table*}

The largest chemical systems that have been computed exactly have 10 or fewer electrons, which is far too few to represent a solid.
Most QMC calculations use the fixed-node (or fixed-phase) diffusion Monte Carlo technique, which requires a trial wave function to set the position of the nodes.
The available wave functions are summarized in Table~\ref{table:trial_wf}.
The Slater-Jastrow (SJ) wave function is by far the most common, and often offers a good compromise between efficiency and accuracy.
While there are no strict rules, the SJ wave function is often less accurate when there is a quasi-degeneracy.
With current algorithms\cite{toulouse07}, it is possible to optimize many parameters in these trial wave functions.
It is important to remember that the variational nature of the technique is key here: all parameters can be optimized to minimize the total energy.

The fixed-node diffusion Monte Carlo method has only a few errors.
There are some, such as the finite simulation cell size and the DMC timestep error, that are controllable using a reasonable amount of computer time.
The only two uncontrolled errors for the ground state are the fixed-node error and the pseudopotential error.
As mentioned before, the fixed-node error is always positive compared to the ground state energy and thus can be examined and minimized by using different trial wave functions. 
The pseudopotentials must be tested carefully against experiment and known results to ensure accuracy.

\subsection{Special considerations for transition metals}

\begin{figure}
\includegraphics[width=\columnwidth]{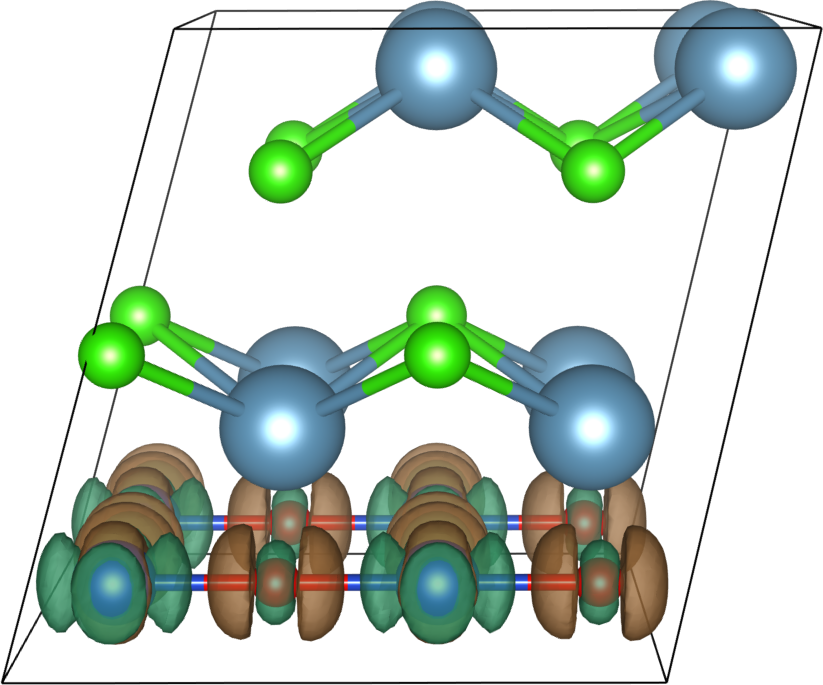}
\caption{The difference in charge density between PBE and PBE0 for Ca$_2$CuO$_2$Cl$_2$. 
Blue isosurfaces represent where PBE has more density while brown represents where PBE0 has more density.}
\label{fig:cuprate_density}
\end{figure}

Many correlated electron systems contain transition metals, in particular the $3d$ transition metals, which have special physics that must be accounted for.
In these systems, the short-range correlation between electrons is very strong because the $3d$ orbitals are quite localized.
The energy of the $3d$ orbitals depends strongly on the quality of the short-range correlation, which then affects the degree of hybridization between the $d$ orbitals and any ligands.
This was first realized in transition metal molecules\cite{wagner_quantum_2003}; it was shown that this effect is stronger than the multideterminant character\cite{wagner_energetics_2007} in those systems.
Later, it was shown\cite{kolorenc_wave_2010,kolorenc_quantum_2008} that this physics applies to transition metal oxide bulk materials in a very similar way to the molecules.

In Fig~\ref{fig:cuprate_density}, we show the difference in density between a Slater determinant formed from DFT(PBE) and DFT(PBE0), a hybrid functional.
Roughly speaking, adding exact exchange to form the hybrid functional effectively reduces the energy of the copper $d_{x^2-y^2}$ state and increases the penalty for double occupation, relative to the oxygen $p$ levels.
This causes the magnetic moment of the copper atom to increase relative to DFT(PBE), and for the remaining charge to settle onto the oxygen $p$ states, which is seen in Fig~\ref{fig:cuprate_density}.
This physics is relevant to 3-band models of the cuprates\cite{kent_combined_2008}, in which the relative positions of the $p$ and $d$ levels are poorly determined by traditional DFT approaches.
As it turns out, the PBE0 orbitals have a lower energy in this system by about 0.4 eV/formula unit, which along with the variational theorem allows the method to determine the best one-particle starting point in the presence of correlation.

\subsection{Implementations}

There are now several good implementations of first-principles quantum Monte Carlo, available to researchers. 
In particular, QMCPACK\cite{_qmcpack_????}, CASINO\cite{_casino_????}, and QWalk\cite{wagner_qwalk:_2009} are full-featured supported codes that can be used for production calculations.

\section{Application to strongly correlated electron systems}

The quantum Monte Carlo methods we have discussed have been around  in nearly their current form for many decades, with the notable exception of efficient energy optimization\cite{toulouse07}; the current algorithms are not so different qualitatively from those in 1971\cite{grimm_monte-carlo_1971}.
However, the implementation of these algorithms has been improved, though faster implementations, improved approximations, and more automatic calculations.
Simultaneously, computers have become many orders of magnitude faster than they were in 1971, which allows the treatment of more particles and potentials with higher variance.
In Table~\ref{table:impact}, we show the progress of first-principles QMC calculations on systems which may be called strongly correlated.
As can be seen, other than a few heroic calculations in the early 2000's, the application of QMC to strongly correlated systems has exploded in the past few years, largely due to the efforts of a few groups as the method has become viable for these systems.

\begin{table}
\caption{Progress of first principles FN-DMC calculations on correlated electron systems.}
\label{table:impact}
\begin{tabular}{l|c|c|}\hline
Material & Year\\
\hline
NiO Mott insulator\cite{needs_diffusion_2003} &  2001 \\
MnO Mott insulator\cite{lee_quantum_2004} &  2004 \\
FeO phase transition\cite{kolorenc_quantum_2008} &  2008 \\
Undoped cuprates\cite{foyevtsova_ab_2014,wagner_effect_2014} & 2014 \\
Doped cuprates\cite{wagner_ground_2015} & 2015 \\
Cerium\cite{devaux_electronic_2015} &  2015 \\
VO$_2$ metal-insulator\cite{zheng_computation_2015} & 2015 \\
Defects in ZnO\cite{santana_structural_2015} & 2015 \\
MnO phase stability\cite{schiller_phase_2015} & 2015 \\
ZnO/ZnSe gaps\cite{yu_systematic_2015} & 2015 \\
NiO\cite{mitra_many-body_2015} & 2015 \\ \hline
\end{tabular}	
\end{table}

\subsection{Quantitative accuracy}

\begin{table*}
\caption{Properties that can be calculated for correlated electron materials using quantum Monte Carlo techniques and typical accuracies. Since the sample size is relatively small, these numbers are very approximate. }
\label{table:accuracy}
\begin{tabular}{l|c|c|}\hline
Property & Typical error (FN-DMC) & Typical error (DFT(PBE)) \\
\hline
Atomization energy &  5\% & 20\% \\
Electronic gap & 10\% & 50-100\% \\
Superexchange constant $J$ & 10\% & 100\% \\
Lattice constant & 1\% & 2\% \\ \hline
\end{tabular}
\end{table*}

\begin{figure}
\includegraphics{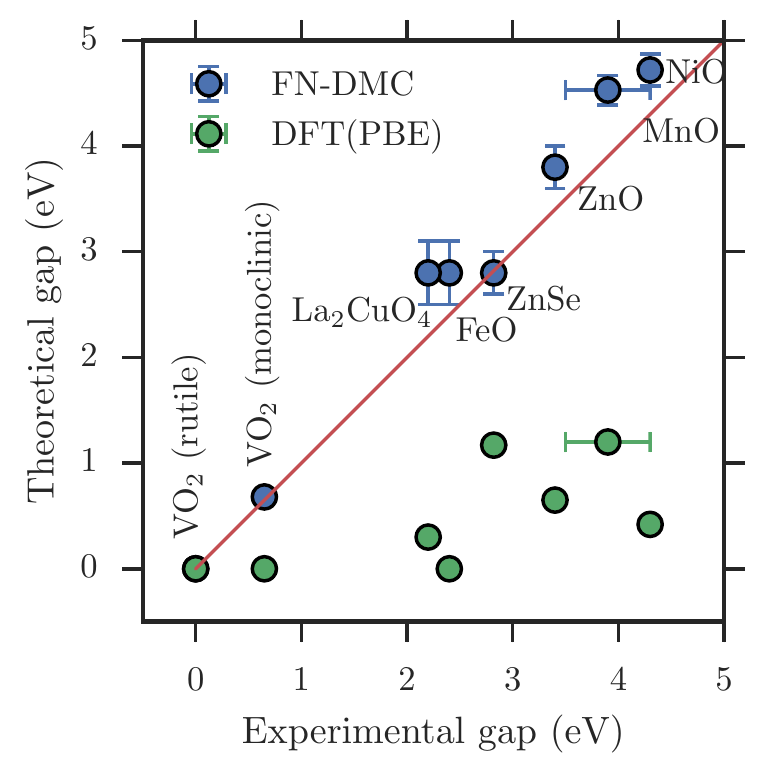}
\caption{Gaps of correlated electron systems calculated by fixed node diffusion Monte Carlo. Data taken from Refs~\cite{zheng_computation_2015,kolorenc_quantum_2008,yu_systematic_2015,schiller_phase_2015,santana_structural_2015,mitra_many-body_2015,wagner_effect_2014}. } 	
\label{fig:gaps}
\end{figure}

One important reason to seek out quantum Monte Carlo is for higher accuracy and fidelity to the real system without the need for adjustable parameters.
For example, the metal-insulator transition in VO$_2$ does not occur in the most commonly-used density functional, PBE\cite{perdew_generalized_1996}, which instead predicts the system to be metallic for both the metallic and insulating structure. 
Hybrid functionals, which should improve the accuracy, do not improve the description\cite{grau-crespo_why_2012}.
Since the basic effect does not appear in these theories, it is not clear how to establish a mechanism,  one of the main motivations for a simulation.
In contrast, the improved accuracy in treating correlations, and in particular the balance between localized states and delocalized states, allows FN-DMC to describe the metal-insulator transition without any corrections.
The mechanism for the metal-insulator transition could thus be ascertained in a clear way\cite{zheng_computation_2015} and specific predictions could be made.
Similar accuracy has been found for a number of real systems (Fig~\ref{fig:gaps}).

Since the first principles many-body wave function is calculated in FN-DMC, the total energy is a meaningful quantity, and since the method attains upper bounds very close to the exact energy, the total energy differences are also quite accurate.
The atomization energies are approximately four times more accurate than DFT numbers for correlated systems\cite{kolorenc_applications_2011,wagner_energetics_2007,bande_vanadium_2008}.
To get an idea of what this increase means for the prediction of new materials, consider that traditional DFT obtains accuracies  from $\pm$ 70 meV/atom\cite{hautier_accuracy_2012} to $\pm$240 meV/atom\cite{lany_semiconductor_2008}, depending on which energy differences are considered.
The difference in enthalpy between phases is typically on the order of 100 meV/atom. 
So, by decreasing the error by a factor of four, the total energy is now more accurate than the energy differences between candidate materials, which decreases both the false positive rate (a structure is predicted stable when it is not), and the false negative rate (a structure is excluded from a search when it is actually stable) by a factor of $e^{-16}$, assuming that the errors are roughly Gaussian distributed.
The disadvantage of the QMC techniques for this problem is that the calculations are computationally expensive and may not be practical for high throughput; however, the next generation of large computational resources may enable this application.

The accuracy in total energy is also useful for excited states: by preparing a nodal surface which restricts the FN-DMC solution to a symmetry different from the ground state, information about excitations can be gleaned.
This has been used to calculate the gaps of both semiconductors and Mott insulators accurately and to calculate the energy difference between different magnetic orderings.
The ability to probe magnetic energetics allows one to compute effective Heisenberg exchange constants $J$, as was done recently for the cuprates\cite{foyevtsova_ab_2014,wagner_effect_2014}.
One can take this further; in Ref~\cite{wagner_effect_2014}, the coupling of the effective Heisenberg system to phonons was also evaluated.
The general theory to use quantum Monte Carlo techniques to derive effective models was recently developed\cite{changlani_density-matrix_2015}, and we give an alternate derivation in Section~\ref{sec:downfolding}.

\subsection{The surprising accuracy of the Slater-Jastrow wave function}

The fact that the Slater-Jastrow wave function can obtain accurate results for these systems is surprising, at least to many who understand quantum Monte Carlo methods well. 
After all, these systems are termed strongly correlated electrons.
In quantum chemistry, a system is called strongly correlated if the ground state wave function is highly multi-determinantal, which the Slater-Jastrow wave function is not.
If strongly correlated materials were highly multi-determinantal, then one would expect that the Slater-Jastrow wave function would not perform as well as it does, as in the case of the molecular systems considered in Sec~\ref{sec:trial_wavefunctions}.
In those systems, there is a dramatic failure of FN-DMC based on a single determinant reference.
However, that does not seem to be the case for these solids, at least so far.
Perhaps part of this effect is due to the fact that a wave function with a Jastrow factor, and the fixed node wave function, include an infinite number of highly relevant determinants in a Slater determinant expansion, even if the expansion is not very flexible.

\subsection{Characterization of the correlated state}

Since QMC methods simulate electron interactions explicitly, if one can think of a correlation function to measure, it can probably be measured.
There are a few, however, that are particularly useful. 

{\bf Static/equal time structure factor.} The static structure factor $S({\bf q})$, the averaged squared modulus of the Fourier transform of the electron density, is related to the Fourier transform of the electron-electron correlation function. 
This is the frequency integral of the charge-charge correlation function measured in X-ray experiments, and it gives information about the density-density fluctuations in the system. 
$S({\bf q})$ can be used for correcting finite size errors in the potential energy\cite{chiesa_finite-size_2006}.

{\bf Reduced density matrices.} 
A reduced density matrix is a projection from the many-body wave function, which is a pure state but in a very high dimension, to a lower dimension. 
This can be done by partitioning space, or by partitioning into one-body (1-RDM), two-body (2-RDM), and so on.
In an uncorrelated theory such as Hartree-Fock, the entire state can be described in terms of the 1-RDM; it is a matrix that has eigenvalues all equal to one or zero.
When correlations are introduced, the 1-RDM now has eigenvalues that are between zero and one and measures the electronic correlation.
For example, the quasiparticle weight in Fermi liquid theory is given by a jump in eigenvalues of the 1-RDM in a homogeneous electron gas, which was recently calculated using quantum Monte Carlo techniques\cite{holzmann_momentum_2011}.
The 1-RDMs are thus a window into effective low-energy theories using accurate solutions. 
We will expand on this in \ref{sec:downfolding}.

In principle, even subtle states like superconductivity and superfluidity can be detected by computing the one or two particle reduced density matrix.
According to Yang\cite{yang_concept_1962}, a condensed state appears when there is off-diagonal long-range order in this density matrix.
While as far as we know, no one has demonstrated this in a realistic fermion system, it has been shown to work in models of ultracold atomic systems\cite{li_atomic_2011} and in liquid $^4$He\cite{C095}.

\subsection{Application to the cuprates}
\begin{figure}
\includegraphics{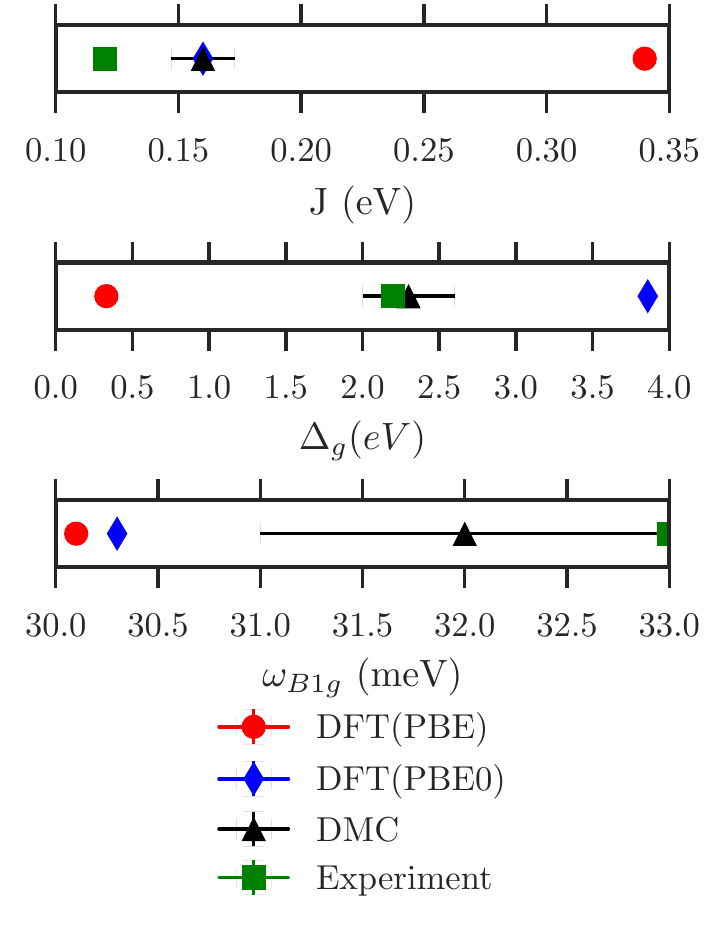}
\caption{Summary of benchmark data for La$_2$CuO$_4$ using several methods, from Ref~\cite{wagner_effect_2014} and references therein.  DMC is able to reproduce experimental values for several parameters that depend on treatment of correlation. We have included a 0.5 eV correction to the DMC values because the gap was calculated for the direct $\Gamma$ point transition, while La$_2$CuO$_4$ has an indirect gap.}
\label{fig:cuprate}
\end{figure}

To make the application to chemical Hamiltonians more concrete, we consider recent work on the electronic structure of the high temperature superconducting cuprate materials\cite{wagner_effect_2014,foyevtsova_ab_2014,wagner_ground_2015}.
For these materials, standard DFT calculations obtain qualitatively incorrect results, with no gap at all in the undoped case, and the doping behavior is qualitatively incorrect, with holes occupying copper $3d$ states instead of the oxygen $2p$ states.
This is because the energy difference between the $2p$ and $3d$ states in standard DFT is incorrect\cite{kent_combined_2008}.
The atomic energy levels are affected significantly by short-range correlations, which are necessary to set the basic physics of the cuprates.
These short range correlations are also evident in the values of the gap and the effective superexchange constant $J$. 

In Fig~\ref{fig:cuprate}, we show the results of performing DMC calculations on La$_2$CuO$_4$, compared to standard DFT.
We have included a hybrid DFT, DFT(PBE0), which mixes 25\% Hartree-Fock exchange into the DFT. 
This hybrid functional helps to correct the self-interaction error present in generalized gradient approximations such as DFT(PBE), at the cost of introducing an additional parameter in the calculation.
Even with this additional parameter, it is not possible to obtain both an accurate value for the superexchange and the gap, with DFT(PBE0) performing well for the magnetic properties, but overestimating the gap significantly.
The higher accuracy attained here is not just about getting the correct quantitative number.
It is a sign that the calculation has successfully described the short-range electron correlations to sufficient accuracy to obtain the values shown.

With the predictive accuracy available with DMC, it was then possible to calculate the ground state of a hole in the cuprates.
This was recently\cite{wagner_ground_2015} done, and a very promising candidate ground state description was obtained. 
While there have been many proposals for the ground state of the hole, this one has extra weight because it is based on a proven accurate method for simulating electron correlation.

\section{Deriving effective models from first principles QMC}
\label{sec:downfolding}
QMC is unique in that it doesn't have a model for the electron correlations; they are simulated directly.
This means that an effective model can be based on an analysis of the QMC solution as we mentioned for $^3$He in Sec. \ref{sec:helium3}.

Let's call the first principles Hilbert space $\hilbert_{fp}$ and Hamiltonian $H_{fp}$. 
The low-energy Hilbert space and Hamiltonian are then $\hilbert_m$ and $H_m$.
Then a good low-energy Hilbert space $\hilbert_m \subseteq \hilbert_{fp}$ satisfies the constraint that 
\begin{align}
\ket{\Psi} \in \hilbert_{fp} \text{ and } \bra{\Psi}H_{fp}\ket{\Psi} \leq E_c \notag \\
\implies \ket{\Psi} \in \hilbert_{m},
  \label{eqn:space_matching}
\end{align}
for a cutoff energy $E_c$.
The second principle necessary is if $\ket{\Psi} \in \hilbert_m$, then 
\begin{equation}
\bra{\Psi} H_m \ket{\Psi} = \bra{\Psi} H_{fp} \ket{\Psi}.
\label{eqn:energy_matching}
\end{equation}
By applying these principles, one can derive a method to calculate $H_m$ from QMC calculations.

Typically, we write $H_m$ in second quantization form; that is, 
  \begin{equation}
 H_m  = E_0 + \sum_{ij} t_{ij} c_i^{\dagger} c_j  + \sum_{ijkl} V_{ijkl} c_i^{\dagger}c_j^{\dagger} c_l c_k  .
  \end{equation}
In the Hilbert space of $H_{fp}$, the $c_i$'s are one-body functions in real space.
This representation allows us to study changes within the small Hilbert space without explicitly considering degrees of freedom that may not change, such as the occupation of core orbitals.
One can thus see how to evaluate the left hand side of the equation in Eqn~\ref{eqn:energy_matching}.
\begin{align*}
\bra{\Psi} H_m \ket{\Psi} &=  E_0 + \sum_{ij} t_{ij} \bra{\Psi} c_i^{\dagger} c_j\ket{\Psi}   \\
& + \sum_{ijkl} V_{ijkl} \bra{\Psi} c_i^{\dagger}c_j^{\dagger} c_l c_k \ket{\Psi} .
\end{align*}
The expectation values are the density matrix elements, which can be evaluated using quantum Monte Carlo techniques\cite{mcmillan_ground_1965,wagner_types_2013}.

There is thus a general algorithm that allows for downfolding:
\begin{enumerate}
\item Sample a large number of many body wave functions $\{\Psi_k\}$
\item Evaluate  $\bra{\Psi_k} H_{fp} \ket{\Psi_k} $, reject any larger than $E_c$.
\item Evaluate $\bra{\Psi_k} c_i^\dagger c_j \ket{\Psi_k}$ for a complete basis $\{c_i\}$.
\item Select a subset of $c_i$'s such that Eqn~\ref{eqn:space_matching} is satisfied.
\item Choose which Hamiltonian parameters $t_{ij}$ and $V_{ijkl}$ to allow to be nonzero.
\item Minimize deviations from Eqn~\ref{eqn:energy_matching} to fit the Hamiltonian parameters.
\end{enumerate}

What is QMC bringing to this?
It is generating wave functions in the low-energy subspace. 
That is, take the spectral expansion of a general wave function $\ket{\Psi}$ in eigenfunctions of the first principles Hamiltonian $\ket{\Phi_i}$, with eigenvalue $E_i$; that is, $\ket{\Psi}=\sum_i c_i \ket{\Phi_i}$.
If $\ket{\Psi}$ contains non-zero coefficients for $\ket{\Phi_i}$ that is not an element of $\hilbert_m$, then the matrix elements are contaminated; that is, $\ket{\Psi}$ is no longer completely in $\hilbert_m$. 
The purpose of the QMC calculation is to remove contamination from high-energy states from the fitting wave functions. 
The procedure is exact when the $\ket{\Psi_k}$'s are completely contained in the small Hilbert space $\hilbert_m$.

This technique is similar in spirit to fitting an effective classical potential to ab-initio calculations, commonly done in preparation for molecular dynamics calculations. 
In particular, it is close to the force-matching technique, in which points are sampled in configuration space and the forces are used as a guide to the effective classical potential.
It is not necessary to fit to the lowest energy configuration to obtain a high quality potential; it is only necessary to sample near enough to the lowest energy configuration that the potential includes it.

\section{The place of first principles QMC in the pantheon of computational techniques}

\paragraph{Accuracy} When dealing with problems in electronic structure it is good to occasionally remember how accurate the calculations need to be, as we touched upon earlier in Section 2.2. Ambient temperature sets a scale for many physical questions, room temperature being equal to 26meV.  Hence to decide which crystal structure or which isomer will be stable one needs to order them to a fraction of this energy, say 10meV. This is at least one order of magnitude more accurate than present day DFT calculations The barrier energy of transition states for reactions is another example of where temperature sets the scale of accuracy.   for many practical purposes, the computational scheme that can deliver this precision in a robust fashion with the available computation resources will be the one that is used. It is not necessary for either DFT or QMC to be ``exact'', only that its error is predictably more accurate than 10meV for these kinds of problems. The ``sign problem'' while very important does not have to be ''solved''; it is enough that the errors in relative energies be robustly less than 10meV.

\paragraph{Scalability.}  The other aspect of this issue is how methods scale with system size.  Typically one finds that both DFT and QMC methods scale asymptotically as N$^3$, though for insulators with a large band gap the exponent can be reduced. In practice the  question should not be how the algorithms scale, but whether one can do a large enough system so that the relative finite-size errors can be made less than 10meV. How many electrons are needed to model a given system? How can finite size errors be corrected for? How large are the computer resources? 

\paragraph{Qualitative vs. quantitative.} Although QMC has the reputation of giving accurate results with a generous input of computational resources, it should also be realized that the whole process of the different formulation of quantum statistical mechanics in terms of a mapping to a stochastic process, leads to a different way of understanding quantum systems, particularly suited to highly correlated ones. An example is in superfluid $^4$He. Feynman made the connection between macroscopic exchange of imaginary time path integrals and the transition to the superfluid.  What came out of the struggle to use this picture to simulate liquid helium, were a theoretical results: the winding number formula to measure and characterize the superfluidity, and the binding formula to measure Bose condensation\cite{C095}. We can expect more such qualitative understanding coming from the stochastic picture that QMC provides  for correlated fermion systems. 

\section{Conclusion}

Quantum Monte Carlo methods offer a way to obtain high fidelity simulations of a many-body quantum system.
By high fidelity, we mean that the simulation has minimal approximations and simulates the many-body system directly, without intermediate models or obfuscations.
These aspects of the QMC methods enable discovery of correlated fermionic systems in two major ways outlined in this Report: high accuracy compared to experimental results on a system and the ability to use the simulation to provide additional data about how correlations between particles lead to observed properties.

The high accuracy of QMC techniques has often been used for systems that are not typically called strongly correlated, in order to provide benchmark data.
In strongly correlated systems, the QMC techniques can provide qualitatively better results than simpler techniques.
In this Report, we summarized the homogeneous electron gas, of which QMC calculations provide benchmark data that make up the foundation of density functional theory.
We also summarized the recent progress in applying QMC techniques to highly realistic models of correlated electron materials.
In system such as the cuprates and VO$_2$, the QMC calculations obtained high accuracy and a clear picture of important physics in these materials.

Explicit simulation of fermion correlations allows for analysis of the many-body physics from the simulation.
In solid $^3$He, the many-body simulation allowed for a {\it discovery} from the simulation of a new way to describe the physics of that system.
We also showed recent developments in formalizing similar types of analysis essentially using data mining of the many-body space.
While these benefits are not unique to QMC, it is often one of few techniques that can treat large enough systems to sufficient accuracy to make the analyses useful.

{\bf Acknowledgements} 
This material is based upon work supported by the U.S. Department of Energy, Office of Science, Office of Advanced Scientific Computing Research, Scientific Discovery through Advanced Computing (SciDAC) program under Award Number DE-SC0008692.

\bibliographystyle{unsrt}

\bibliography{dmc,bib-dmc,sces_ropp}

\end{document}